\documentclass[proceedings]{rmaa}

\title {Temperature Structure and Chemical Abundances in Gaseous Nebulae}
\author{Manuel Peimbert
	\affil{Instituto de Astronom\'{\i}a, Universidad Nacional Aut\'onoma de
	M\'exico}}
\fulladdresses{
\item M. Peimbert: Instituto de Astronom\'{\i}a, UNAM,
Apartado Postal 70-264, Cd. Universitaria, C.P. 04510, M\'exico, D.F., M\'exico
(peimbert@astroscu.unam.mx).}
\shortauthor{Peimbert}
\shorttitle{Temperature \& Abundances in Gaseous Nebulae}


\resumen{Discuto algunos de los resultados presentados en este simposio sobre:
nebulosas planetarias, regiones H~{\sc{ii}}, evoluci\'on qu\'{\i}mica de
galaxias y la determinaci\'on de la abundancia primordial de helio. Para tener
una visi\'on m\'as general de este simposio recomiendo leer todos los
art\'{\i}culos incluidos en este volumen.}

\abstract{In this summary I review some of the results presented in this
symposium relative to: planetary nebulae, H~{\sc{ii}} regions, chemical
evolution of galaxies, and the determination of the primordial helium
abundance.  To get a more general perspective of this symposium I encourage you
to read all the contributions to these proceedings.}

\begin{document}
\maketitle
\section{Overview}

To produce an accurate model of a gaseous nebula we should take its most
relevant properties into account. These properties include: the geometry, the
temperature structure, the density structure, the velocity structure, the dust
content, the chemical abundances (together with possible chemical
inhomogeneities inside a given object), and the energy sources.

Precise models of individual nebulae permit to determine accurate abundances
and the abundances permit to test models of stellar evolution, galactic
chemical evolution, and the evolution of the Universe as a whole.

To have a good model of a given gaseous nebula a very good knowledge of the
temperature structure is needed. The temperature structure is crucial for the
determination of accurate chemical abundances. For the best observed objects
usually we have a value of the average temperature $T_o$ and of the mean square
temperature fluctuation, $t^2$. When $t^2$ agrees with the value predicted by
chemically homogeneous photoionization models we are confident of the derived
chemical abundances. Often the observed values of $t^2$ are higher than those
predicted by the models and a source for the discrepancy should be sought. Two
points should be made here: $a$) the observational errors present in the $t^2$
determinations are high, but maybe lower than I expect because the overwhelming
majority of the observational $t^2$ values present in the literature are
positive, b) as Daniel P\'equignot mentioned during his talk $t^2$ is just an
empirical parameter that should be adjusted by the model, a larger
observational value for $t^2$ than that predicted by the model is only telling
us that something is wrong with the model, but it is not telling us what is
wrong nor which is the temperature structure. This review will be mainly
centered on the relevance of the temperature structure in the determination of
the chemical composition of different objects, another view of the role of the
electron temperature in abundance determinations is presented by Stasi\'nska
(2001).

\section{Planetary Nebulae}

The abundances derived from permitted lines run from similar to about an order
of magnitude higher than those derived from forbidden lines (see the review by
Liu 2001). By assuming that collisional deexcitation is not important (low
density limit) and that the objects are chemically homogeneous it is possible
to reach agreement between both types of determinations adopting a $t^2$ $>$
0.00.

The $t^2$ values determined from observations are in the 0.00 to 0.09 range,
while those values predicted by chemically homogeneous photoionization models,
CHPM, are in the 0.005 to 0.025 range. We can divide the well observed PNe in
three groups: $a$) those that have $t^2$ values smaller than 0.025, they can be
fitted with CHPM and comprise about a third of the well studied cases, $b$)
those with intermediate $t^2$ values, in the 0.025 to 0.045 range, and c) those
with $t^2$ larger than 0.045, most of these objects are of Type I (Peimbert
1978; Peimbert et al. 1995) and show complex velocity fields reaching velocity
differences of many hundreds of km s${^{-1}}$, for these objects the
deposition of mechanical energy might be significative. A lot of effort 
has been put into the determination of $t^2$ and special attention has been given 
to those objects with the largest $t^2$ values.

To explain the $t^2$ differences between the predicted values from CHPM and the
observed values at least eight possible causes have been suggested in the
literature (see the review by Esteban 2001): $a$) large density variations,
$b$) chemical inhomogeneities, $c$) deposition of mechanical energy due to
shocks or dissipation of turbulent motions, $d$) enhanced dielectronic
recombination (Garnett $\&$ Dinerstein 2001), $e$) shadowed regions ionized by
indirect radiation from the nebula rather than direct radiation from the
ionizing star (Mathis 1995), $f$) magnetic reconnection (Ferland 2001), $g$)
observational errors, and $h$) errors in the atomic parameters.

Some of these causes might be present in some objects and not in others. Only a
careful analysis of a given object will indicate the relative importance of
each of them.

One question that we want to answer is: which are the representative abundances
for the whole nebula, those provided by the forbidden lines or those provided
by the recombination lines? The answer is fundamental to constrain the
evolution of intermediate mass stars and the chemical evolution of the
Galaxy. If the effects due to $b$) and $d$) dominate then the representative
abundances for the bulk of the mass ejected are those given by forbidden lines
(see Liu et al. 2000; Liu 2001; P\'equignot et al. 2001), alternatively if
effects due to $a$), $c$), and $e$) dominate then the representative abundances
are those given by the permitted lines. Carigi (2001) has constructed models of
the chemical evolution of the Galaxy based on observational yields of carbon
derived from recombination lines and from forbidden lines of planetary nebulae,
she finds that the models that use the yields based on permitted lines agree
better with the observational constrains provided by H~{\sc{ii}} regions and
stars of the solar vicinity, than the models based on the yields derived from
forbidden lines.

\section{Galactic and Extragalactic H~{\sc{ii}} Regions}

There are two different problems related with the temperature structure 
of H~{\sc{ii}}  
regions that are still controversial: $a$) typical observed $t^2$ values are
in the 0.01 to 0.04 range, while typical values predicted by CHPM's are in the
0.005 to 0.020 range. The differences in $t^2$ between CHPM's and observations
of H~{\sc{ii}}  regions are smaller than in PNe but I think that they are real (see the
review of Esteban 2001), and $b$) in general photoionization models predict
$T$(O~{\sc{iii}}) values smaller than observed (Stasi\'nska \& Schaerer 1999;
Luridiana, Peimbert, \& Leitherer 1999; Luridiana \& Peimbert 2001; Luridiana,
Peimbert, \& Peimbert 2001; Rela\~no, Peimbert, \& Beckman 2001) indicating the
possible presence of an additional heating source not considered by the models.

The warning mentioned by Viegas (2001) regarding point $b$) above should be
considered: models depend on many assumptions and a very good model is needed
before we accept its implications. For example the difference between the
observed and predicted $T$(O~{\sc{iii}}) values depends on the adopted filling
factor, $\epsilon$; the difference between the observed and predicted
$T$(O~{\sc{iii}}) values for I~Zw~18 disappears for models with values of
$\epsilon > 0.3$.

\section{Chemical Evolution of Galaxies}

One of the controversial issues in the study of the chemical evolution of
irregular galaxies is the low effective yield of oxygen derived from
observations, one solution to this problem is to assume the presence of O-rich
galactic outflows.

Several lines of reasoning indicate that O-rich outflows produced by gas rich
irregular galaxies are unlikely. Larsen, Sommer-Larsen, \& Pagel (2001) from
chemical evolution models find that the N/O versus O/H relationship indicates
that that O-rich outflows have not played an important role in nearby irregular
galaxies (redshifts $\sim 0$). A similar result has been obtained by Carigi et
al. (1995) and Carigi, Col\'{\i}n, \& Peimbert (1999) based on the C/O versus
O/H values for nearby irregular galaxies. Tenorio-Tagle (2001) from the mixing
of metals argues that it is difficult to expel gas in dwarf irregulars with a
low rate of star formation, moreover the mass lost would be of well mixed
material; in this context Silich et al. (2001) analyze VII~Zw~403, a metal poor
irregular galaxy, and conclude that the heavy elements produced during the
present starburst will not be ejected into the interstellar medium. One way to
reduce the difference between the observed effective yield for oxygen and the
yield predicted by models is the presence of dark matter (e.g. Carigi et
al. 1999).

The N/O versus O/H relation has been studied by many authors (e.g. Garnett
1990, 2001; Pagel 1992; Shields 2001; Skillman 2001; Larsen et al. 2001) there 
are some aspects of this relation that need further consideration.

The N/O ratio depends on the temperature adopted, specially for objects with
low electron temperature, therefore the errors in the N/O determinations might
be larger for objects with lower temperatures. To determine the N/O ratio often
the temperature derived from the $\lambda\lambda$ 4363/5007 ratio,
$T$(O~{\sc{iii}}), is used as representative of the O$^+$ and N$^+$ zones; from
photoionization models it is found that for objects with $T$(O~{\sc{iii}}) $>$
12360K, the temperature of the O$^{++}$ region is higher than the temperature
of the O$^+$ region, the opposite is found for objects with $T$(O~{\sc{iii}})
$<$ 12360K (e.g. Stasi\'nska 1990); if this effect is not taken into account the
N/O value for metal rich H~{\sc{ii}} regions (those with $T$(O~{\sc{iii}}) $<$
12360K) will be underestimated while for metal poor H~{\sc{ii}} regions (those
with $T$(O~{\sc{iii}}) $>$ 12360K) N/O will be overestimated.

Often it is assumed that the N/O ratio is equal to the N$^+$/O$^+$ ratio
(assuming that the O$^+$ zone coincides with the N$^+$ one), while some
photoionization models indicate that this is the case, others indicate that it
is at best a fair approximation Rela\~no et~al. 2001.

Apparently there are environmental effects present in the N/O versus O/H
relation, while Peimbert \& Torres-Peimbert (1992) find an underabundance of
the N/O ratio for a given O/H ratio in the Bootes Void galaxies, V\'{\i}lchez
\& Iglesias-P\'aramo (2001) find that there is an overabundance of N/O for a
given O/H in the dwarf galaxies of the Virgo cluster.

\section{Primordial Helium Abundance}

The determination of the pregalactic, or primordial, helium abundance by mass
$Y_p$ is paramount for the study of cosmology, the physics of elementary
particles, and the chemical evolution of galaxies (e.g. Boesgaard \&
Steigman 1985; Fields \& Olive 1998;
Izotov et~al. 1999; Peimbert \& Torres-Peimbert 1999; Olive \& Skillman 2000
and references therein).

We will call $Y_p$(nHc) those $Y_p$ values in the literature derived under the
assumption of no contribution to the hydrogen Balmer lines due to collisional
excitation.

The best determinations of $Y_p$(nHc) in the literature are those of Izotov \&
Thuan (1998); Izotov et~al. (1999); and Peimbert, Peimbert, \& Ruiz (2000) that
amount to $0.2443 \pm 0.0015, 0.2452 \pm 0.0015$, and $0. 2345 \pm 0.0026$
respectively. These determinations are based on 45, 2, and 1 extragalactic
H~{\sc{ii}} regions respectively and the differences between the first two and
the last one amount to at least 3$\sigma$.

To study the source of this discrepancy Peimbert \& Peimbert (2001) and
Peimbert, Peimbert, \& Luridiana (2001) decided to compute $Y_p$(nHc) based on
the data by Izotov \& Thuan (1998) and Izotov et~al. (1999). From two different
subsamples of the best observed objects, comprising 12 and 5 objects, found
that $Y_p$(nHc) amounts to $0.2371 \pm 0.0015$ and $0.2360 \pm 0.0025$
respectively. These results are in good agreement with the value derived by
Peimbert et al. (2000) and are significantly smaller than the
values derived by Izotov \& Thuan (1998) and Izotov et~al. (1999).

The main source of the discrepancy between both groups of authors is due to the
treatment of the temperature structure inside the nebulae; while Izotov \&
Thuan (1998) and Izotov et~al. (1999) adopt $T$(O~{\sc{iii}}) to derive the
helium abundance, Peimbert \& Peimbert (2001) and Peimbert et al. (2001) 
from the He~{\sc{i}} line intensities and adopting $t^2 >
0.00$ determine $T$(He~{\sc{ii}}) values 6-11$\%$ smaller than
$T$(O~{\sc{iii}}).  In the self-consistent solutions the smaller
$T$(He~{\sc{ii}}) values imply higher densities; the higher the density the
higher the collisional contribution to the He~{\sc{i}} line intensities and,
consequently, the lower the helium abundances.
 
The baryon energy density, $\Omega_b$, values derived by Peimbert \& Peimbert
(2001) and Peimbert et al. (2001) from the $Y_p$(nHc) values
are significantly smaller than the $\Omega_b$ value derived from the D$_p$
determination by O'Meara et~al. (2001). Before we conclude that a non-standard
big bang nucleosynthesis model is needed to reconcile the differences it is
necessary to analyze further two possible systematic effects: $a$) the
ionization structure of the H~{\sc{ii}} regions, and $b$) the collisional
excitation of the hydrogen lines.

To determine very accurate He/H values of a given H~{\sc{ii}} region we need to
consider its ionization structure.  The total He/H value is given by:

\begin{eqnarray}
\frac{N ({\rm He})}{N ({\rm H})} & = &
\frac {\int{N_e N({\rm He}^0) dV} + \int{N_e N({\rm He}^+) dV} + 
\int{N_e N({\rm He}^{++})dV}}
{\int{N_e N({\rm H}^0) dV} + \int{N_e N({\rm H}^+) dV}},
						\nonumber \\
& = & ICF({\rm He})
\frac {\int{N_e N({\rm He}^+) dV} + \int{N_e N({\rm He}^{++}) dV}}
{\int{N_e N({\rm H}^+) dV}}
\label{eICF}
.\end{eqnarray}

For objects of low degree of ionization it is necessary to consider the
presence of He$^0$ inside the H$^+$ zone, while for objects of high degree of
ionization it is necessary to consider the possible presence of H$^0$ inside
the He$^+$ zone. For objects of low degree of ionization $ICF({\rm He})$ might
be larger than 1.00, while for objects of high degree of ionization $ICF({\rm
He})$ might be smaller than 1.00. The deviations from unity in the $ICF({\rm
He})$ value occur in and near the ionization boundary of a given H~{\sc{ii}}
region, therefore those H~{\sc{ii}} regions that are density bounded in all
directions have an $ICF({\rm He})$ = 1.00. The $ICF({\rm He})$ problem has been
discussed by many authors (e.g. Shields 1974; Stasi\'nska 1983; Pe\~na 1986;
V\'{\i}lchez \& Pagel 1988; Pagel et~al. 1992; Armour et~al. 1999; Peimbert \&
Peimbert 2000; Viegas, Gruenwald, \& Steigman 2000; Viegas \& Gruenwald 2000;
Ballantyne, Ferland, \& Martin 2000; Sauer \& Jedamzik 2001).

Based on models of metal poor H~{\sc{ii}} regions Luridiana et~al. (2001) find
that the $ICF$(He) for some of the best observed objects is very close to 1.00
and consequently that the main difference between the $\Omega_b$ value derived
from $Y_p$(nHc) and D$_p$ is not due to the $ICF$(He). Rela\~no et~al. (2001)
from the spectral types of the ionizing stars of NGC 346 find that about half
of the ionizing photons escape the nebula favoring an $ICF$(He) = 1.00, this
result is also supported by the fit of the lines of low degree of ionization by
their photoionization model. From the work by Zurita, Rozas, \& Beckman (2000)
on the ionization of the diffuse interstellar medium in external galaxies it is
expected that a large fraction of the ionizing photons escapes from the most
luminous H~{\sc{ii}} regions, which favors the assumption that the $ICF$(He) is
very close to 1.00.

Davidson \& Kinman (1985) were the first to estimate the collisional
contribution to the Balmer lines and its effect on the determination of $Y_p$;
they made a crude estimate for I~Zw~18 and concluded that the collisional
contribution to $I$(H$\alpha$) $may$ be roughly 2\%. All the subsequent
determinations of $Y_p$ in the literature have been derived under the
assumption of no contribution to the hydrogen Balmer lines due to collisional
excitation, I have referred to these determinations in this paper as
$Y_p$(nHc).

Notice that to a very good approximation the collisional excitation of the 
Balmer lines does not affect the maximum likelihood method determinations 
of $N_e$(He~{\sc{ii}}), $T$(He~{\sc{ii}})\, $\tau(3889)$, and 
$T$(O~{\sc{iii}}).

{From} a series of {\sc Cloudy} models it is found that the collisional
contribution to $I$(H$\beta$) for I~Zw~18 and SBS~0335-052 is in the 2\% to 6\%
range, for H 29 and NGC~2363 in the 1\% to 2\% range, and for NGC~346 in the
0.6\% to 1.2\% range. Our preliminary results indicate that the primordial
helium abundance including hydrogen collisions, $Y_p$(+Hc), is about 0.0050
larger than $Y_p$(nHc). This problem together with the {\sc Cloudy} models for
I~Zw~18, SBS~0335-052, and H 29 will be discussed elsewhere Luridiana
et~al. (2001). The {\sc Cloudy} models for NGC~2363 and NGC~346 are those by
Luridiana et~al. (1999) and Rela\~no et~al. (2001), respectively.

\bigskip

I should like to express my thanks to Jose Franco, the members of the
Scientific Organizing Committee, and the members of the Local Organizing
Committee for the idea of holding this symposium in honor of Silvia and me. I
am also grateful to all the participants for a very stimulating and enjoyable
meeting.

\end{document}